\documentclass[12pt]{article}
\usepackage{epsfig}

\title{Stochastic  dynamics and  Fokker-Planck equation in accelerator physics}

\author{H. Mais, M.P. Zorzano}

\date{ 
DESY 98-173 November 1998 \\ physics/9901010}

\begin{document}
\maketitle

\begin{abstract}
The aim of this contribution is to study the particle dynamics
in a storage ring under the influence of noise. Some simplified
stochastic beam dynamics problems are treated by solving the
corresponding Fokker-Planck equations numerically.
\end{abstract}

\section{Introduction}

Colliders have become an important tool in high energy physics. 
For example, HERA, the electron-proton collider at DESY, consists of a
820 GeV proton ring and a 30 GeV electron ring. At the intersections of these
two rings the colliding beam experiments H1 and ZEUS are located. 
Besides these colliding beam experiments there are also two internal
target experiments: HERA-B probing the proton beam halo with a wire
target and HERMES using the longitudinally polarized electron beam.
 The experiments require optimal performance of the
collider  i.e. maximum luminosity (collision rates),
high polarization degree and controled beam halo. In order to achieve these
requirements one needs a good  understanding
of all  phenomena and effects which cause a degradation 
 of the beam quality. The beam constitutes
a complicated many particle system of about $10^{13}$ charged, 
ultrarelativistic ($v \approx c$) particles with spin
which are
distributed in 180 bunches. This ensemble is subject to external
electromagnetic fields (dipoles, quadrupoles, multipoles and rf fields), to
space charge fields and wakefields. Furthermore, various scattering mechanisms
(restgas, intrabeam) must be taken into account, and in the case
of the
electrons, radiation phenomena must be included. In addition to these 
effects there are also various sources of noise such as rf noise,
power supply noise, random ground motion and quantum
fluctuations due to the radiation. Altogether, the beam constitutes a 
complicated nonlinear, explicitly stochastic many particle system. 

The goal of accelerator physics is to describe and understand
via suitable models the dynamical behaviour of such a system - in the
ideal case in terms of macroscopic dynamical variables such as particle
density  and polarization density
(in phase space or configuration space).
Mathematically, the system can be described via a stochastic
Liouville equation (for a recent discussion of this approach in
accelerator physics see \cite{Ellison1}) or via a stochastic differential
equation.
 
In the following we will restrict our considerations  to 
the latter case, the Langevin-like description of  dynamical systems.
For more information about
the conceptual foundations of these statistical dynamics problems
we refer the reader to \cite{Ruggiero,Balescu}.

Our paper is organized as follows: In section 2 we summarize some basic facts
about stochastic beam dynamics in accelerators and we remind the reader
of some mathematical results concerning stochastic
differential equations and the Fokker-Planck equation. In section 3
we apply the Fokker-Planck description to certain (simplified)
accelerator problems and models such as beam-beam interaction, rf noise
and diffusion out of a stable rf-bucket. Section 4 consists of a summary
and a list of open problems for future work.
 
\section{Stochastic dynamics}

In the classical approximation and for the problems to be studied below,
the particle dynamics in accelerators 
 can  be written in the
form of a multiplicative stochastic differential equation with
a Gaussian white noise vector process $\xi(t)$
\begin{eqnarray}\label{dyn}
\frac{d}{dt} y(t) = f(y,t) + T(y,t) \xi(t)
\end{eqnarray}
 where $y(t)= (x(t), \eta(t))$   consists
of the n-dimensional phase space vector $x(t)$ 
(n=2,4 or 6 for the orbital motion and n=8 for the spin-orbit
motion \cite{MaisRipken})
 and an
m-dimensional Ornstein-
Uhlenbeck type stochastic process $\eta(t)$,
$f(y,t)$ is an (n+m)-dimensional known
vector function and $T(y,t)$ is an 
(n+m)$\times$(n+m) matrix   (for more details see for example
the review  \cite{Mais}).
In the case of protons,  equation (\ref{dyn})
describes the  stochastic Hamiltonian dynamics of the 
coupled synchro-betatron oscillations, and in the case of electrons -
because of radiation phenomena (radiation
damping and quantum excitation) - (\ref{dyn}) describes a
stochastically and dissipatively perturbed Hamiltonian system \cite{Jowett}.

 With the Ito and Stratonovich 
calculus one has the
mathematical tools to study these multiplicative 
stochastic differential equations, whose solutions are in
general Markovian diffusion processes \cite{Arnold,vKampen,Gardiner,Gard,Horsthemke}. 
In this case, instead of studying the
stochastic differential equations (\ref{dyn}) directly 
\cite{Kloeden,Pauluhn}, one can also study the 
corresponding Fokker-Planck equation. The Fokker-Planck equation
is a partial differential equation for the probability density 
$p(y,t)$ and
the transition density $p(y,t|y_{0},t_{0})$ of the  stochastic process 
 defined by equation (\ref{dyn}), and 
in the
Ito interpretation of this equation  it takes the following form
\begin{eqnarray}\label{FP}  
\lefteqn{ \frac{\partial}{\partial{t}} p(y,t)=     
  -\sum_{i}\; \frac{\partial}{\partial{y_{i}}}[f_{i}(y,t)
  \cdot p(y,t)]+ }  \nonumber \\
 &   &   + \frac{1}{2} \cdot
  \sum_{i,j}\frac{\partial}{\partial{y_{i}}} \cdot \frac{\partial}
   {\partial{y_{j}}}[ \{T(y,t) T^{T}
   (y,t) \}_{ij} \cdot p(y,t)].
\end{eqnarray}
If one integrates $p(y,t)$ over the Ornstein-Uhlenbeck-type
variables  $\eta$, one obtains the probability 
$\bar{p}(x,t) = \int p(x,\eta,t)d\eta$
of finding the system
at time $t$ between $x$ and $x+dx$ in phase space.
Using 
$
N \bar{p}(x,t) dx = dn(x,t)  
$
where $N$ is the total number of particles and where $dn(x,t)$
denotes the number of particles in the  volume element
 $dx$, $\bar{p}(x,t)$  can be interpreted 
(up to a constant) as phase space density of  system (\ref{dyn}).

(\ref{FP}) is a partial differential 
equation with (n+m+1) independent variables and
a solution of this equation requires 
initial conditions and suitable boundary conditions
(natural ($\pm \infty$), periodic, reflecting or absorbing). Only
few exact analytical solutions are available (mainly for the
low (1+1)-dimensional  case). A detailed and 
comprehensive review of the Fokker-Planck equation and a list of
methods for its solution is given in \cite{Risken}
to which  the reader is referred.

In the following we will concentrate on the numerical solution of this
equation by direct discretization of space and time.
The structure of the Fokker-Planck equation suggests an
operator splitting algorithm  \cite{Richtm,Zorzano}
which  will be
applied in the next section to study some
simplified (lower dimensional) stochastic problems and models
in colliders.
\begin{figure}[htb]
\begin{center}
%\parbox[t]{10cm}{
\epsfig{figure=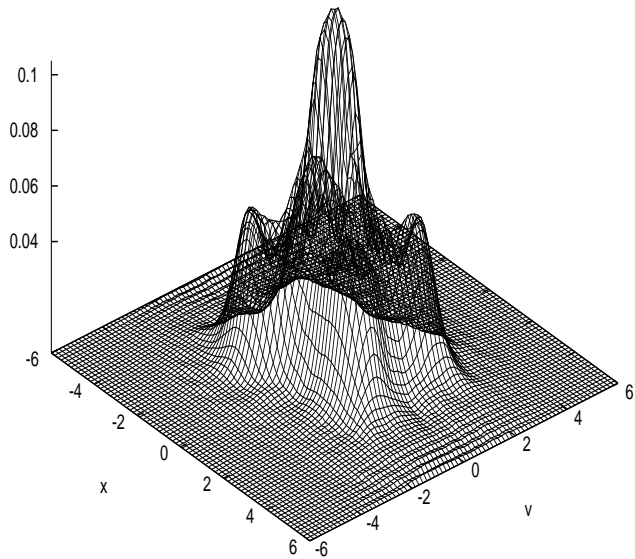,height=10cm,width=10cm}
\caption{\label{BBII}Density distribution, $Q_{x}=0.7$, near fourth order
resonance}
\end{center}
\end{figure}

\begin{figure}[htb]
\begin{center}
\parbox[t]{8cm}{
\epsfig{figure=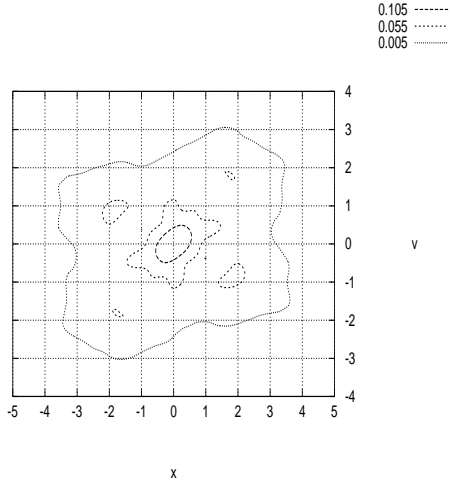,height=8cm,width=8cm}
\caption{\label{BBc}Contour plot of density distribution}}
\end{center}
\end{figure}
\section{Models of stochastic beam dynamics in accelerators}

In this section we consider three
stochastic problems in storage rings. As a first example we study
how an electron is influenced by 
the strong nonlinear fields of a  counter rotating
particle  bunch (weak-strong beam-beam interaction model \cite{Kheifets,Pauluhn,
Gerasimov}). In this case, the equation of motion
for the horizontal betatron oscillations is
given by the following stochastic differential equation
\begin{eqnarray}
\ddot{x} + \tau \dot{x} + Q_{x}^{2} x + f(x,t) = \sqrt{2D} \xi(t) 
 \nonumber
\end{eqnarray}
where $\tau$ is the radiation damping time, $Q_{x}$ is the 
horizontal tune,
$ 
f(x,t)= 8 \pi \zeta_{bb}  \cdot
  \frac{1-\exp(-\frac{x^{2}}{2})}{x} \cdot \delta_{p}(t)
$ is the beam-beam force
with the beam-beam parameter $\zeta_{bb}$, and $\xi(t)$ is the 
Gaussian white noise process of strength $D$. $\delta_{p}(t)$
denotes a strongly localized periodic function.

The numerical solution of the corresponding Fokker-Planck equation
near a fourth order resonance
is depicted in Fig.~\ref{BBII}, and Fig.~\ref{BBc}  shows a contour plot of the density,
which shows  the nonlinear resonance characteristics of the
underlying Hamiltonian dynamics \cite{Izraelev}. 
As initial conditions we have used
a Gaussian distribution localized at the origin
of the two-dimensional phase space ($x, \dot{x}=v$). We also want to mention 
that this model has been used to compare various 
numerical tools to study stochastic systems such as cell-mapping methods,
Monte-Carlo methods and finite differences \cite{www}.

An important problem in proton storage rings is 
to study the influence of rf
noise on the particle stability (see for example \cite{Dome,Shih,Krinsky,
Pauluhn}). Here, we investigate the diffusion out of a stable
rf bucket under the influence of random energy losses (for example
due to scattering) and under Gaussian white noise. The dynamics is
governed by
\begin{eqnarray}
\ddot{\phi} + V_{1} \sin(\phi) + V_{4} \sin(4 \phi) + \Delta_{p}(t)
 = \sqrt{2D} \xi(t)   \nonumber
\end{eqnarray}
where $V_{1}, V_{4}$ denote the voltages  of a two-rf system, and 
where $\Delta_{p}(t)$ denotes the average energy loss due to
scattering.
\begin{figure}[htb]
\begin{center}
\epsfig{figure=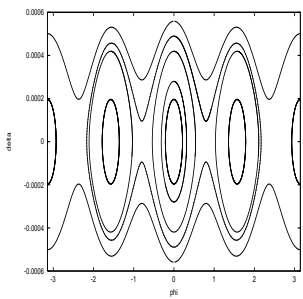,height=4cm,width=4cm}
\caption{\label{fig1}Deterministic longitudinal phase space and bucket structure}
\end{center}
\end{figure}

\begin{figure}[htb]
\begin{center}
\epsfig{figure=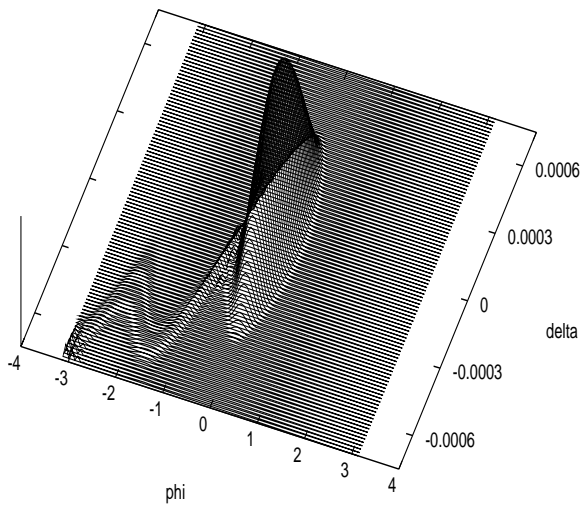,height=10cm,width=10cm}
\caption{\label{fig4}Density after 80000 turns, strong kick $10^{-5}$ every N=100 turns, and strong noise $D=1e-13$.}
\end{center}
\end{figure}

\begin{figure}[htb]
\begin{center}
\epsfig{figure=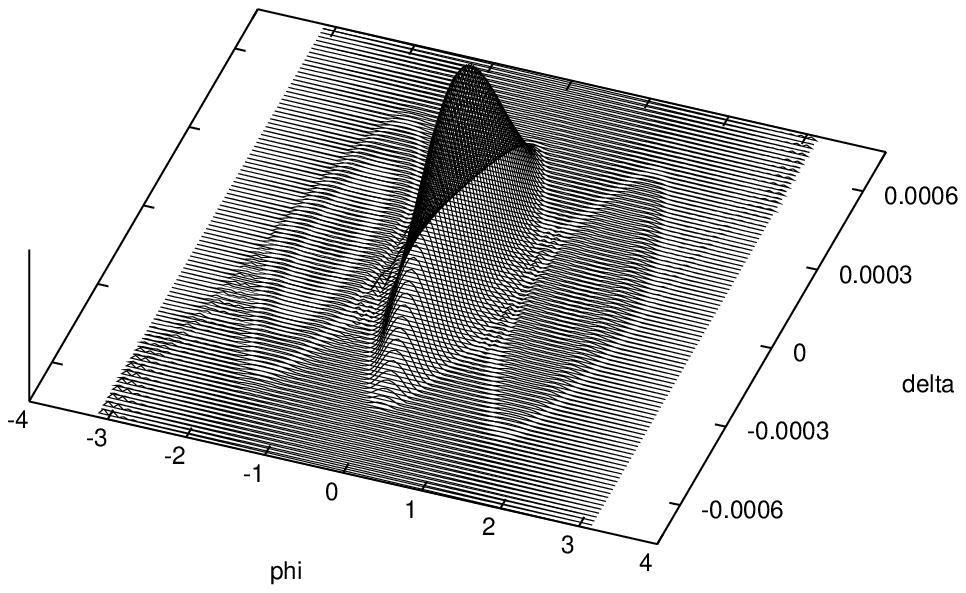,height=10cm,width=10cm}
\caption{\label{fig5}Density after 800000 turns, weak kick $10^{-6}$ every N=100 turns, and weak noise $D=1e-14$.}
\end{center}
\end{figure}

The unperturbed bucket structure is shown in Fig.~\ref{fig1}, and Fig.~\ref{fig4} 
and Fig.~\ref{fig5}
show how the distribution  diffuses outward in phase space thus
forming a coasting beam. The initial distribution was Gaussian and was
localized near the origin.

\begin{figure}[htb]
\begin{center}
\parbox[t]{6cm}{
\epsfig{figure=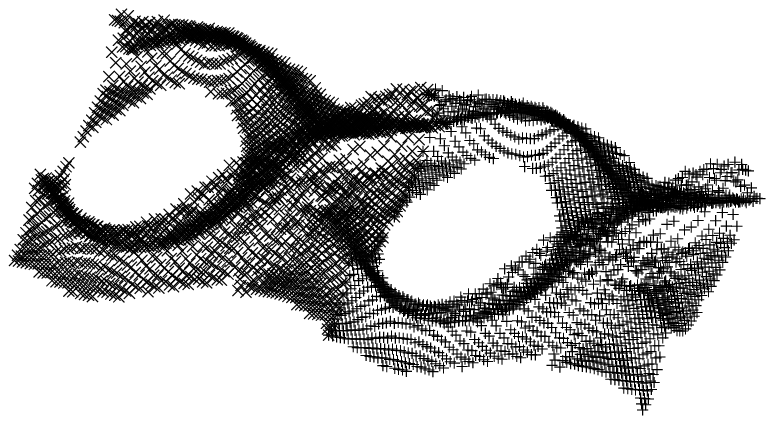,height=6cm,width=6cm}
\caption{\label{f4} longitudinal phase space distribution in case of
coloured amplitude noise integrated over Ornstein-Uhlenbeck variable}}
\end{center}
\end{figure}

The final example we want to show treats the 
longitudinal particle motion  under the influence of
coloured  rf (amplitude) noise with the equation of motion
for the phase 
\begin{eqnarray}
\ddot{\phi} + \Omega_{s}^{2}(1 + \eta(t))\sin(\phi) = 0 \nonumber
\end{eqnarray}
where $\eta(t)$ denotes an Ornstein-Uhlenbeck process
\begin{eqnarray}
\dot{\eta} = - a \eta + \sqrt{2D} \xi(t).   \nonumber
\end{eqnarray}
$D$ and $a$ are parameters which define the correlation time of the
process.
The numerical solution of the corresponding Fokker-Planck equation
integrated over the Ornstein-Uhlenbeck variable ($\eta$) 
is depicted in Fig~\ref{f4} for an initial distribution localized
near the unperturbed separatrix of the system.

More examples which illustrate the usefulness of the Fokker-Planck
description of stochastic problems in accelerators can be found in
\cite{Zorzano}

\section{Summary and discussion}

In this contribution we have shown that stochastic beam dynamics is an
important issue in accelerator physics. Usually 
- via various diffusion mechanisms - noise can lead to a degradation
of the beam quality in a collider (emittance growth, reduced 
lifetime etc). However, the application of cleverly chosen noise
in the transverse plane or longitudinal phase plane can also help to
shape and control the beam and its halo \cite{Sen,Bazzani}. 
Furthermore, we have shown that the Fokker-Planck equation is a suitable
and helpful
mathematical tool to treat these stochastic systems. Since only 
few exact results are available for this partial differential equation
(especially in higher dimensions) one needs a powerful, reliable,
accurate and fast numerical solver. Such a solver
which is based on the operator splitting method has been
developed, and it has been used to study stochastic beam dynamics
problems in accelerators. However, longtime calculations and
higher dimensional problems such as stochastic spin-orbit motion in
realistic colliders with at least (8+1) independent
variables are certainly beyond the capacity of this
and other codes even if parallel algorithms and high
performance computers are used. So -in addition- to these 
numerical studies of the Fokker-Planck equation one also
needs  perturbative methods such as averaging with all its
mathematical subtleties. 
Furthermore, complementary studies of the dynamics via direct
analysis of the underlying stochastic differential
equations  or discrete stochastic maps are important.
Another alternative could be the use of analogue computers
(see the recent interesting review \cite{analogue}).
In addition to these topics noise induced transitions
(stabilization by noise) \cite{Horsthemke} and stochastic resonance
\cite{Jung}
might play an important role in accelerator physics. First
steps 
to investigate these phenomena in storage rings have been undertaken in
\cite{Zorzano}

\section{Acknowledgments}

The authors want to thank V. Balandin, A. Bazzani,
G. Dattoli, J.A. Ellison,
T. Sen, G. Turchetti and L. Vazquez for many stimulating and
helpful discussions. One of us (M.P.Z.) was supported by a 
DESY PhD scholarship and Human Capital and Mobility Contract
Nr. ERBCHRXCT940480.


\begin{thebibliography}{0}
\bibitem{Ellison1}
J.A. Ellison "Accelerators and probability: The special effect of noise in
beam dynamics" in Proc. "Nonlinear and Stochastic Beam Dynamics in
Accelerators - a Challenge to Theoretical and Computational Physics",
L\"{u}neburg 1997, DESY 97-161
\bibitem{Ruggiero}
L.A. Radicati, E. Picasso, F. Ruggiero "Considerations on the
statistical description of charged-beam plasmas" in
"Nonlinear dynamics aspects of particle accelerators" Springer (1986)
\bibitem{Balescu}    
R. Balescu "Statistical dynamics - matter out of equilibrium"
Imperial College Press (1997) 
\bibitem{Jowett} 
J.M. Jowett "Electron dynamics with radiation and nonlinear
wigglers" CERN Accelerator School, Oxford 1985, CERN 87-03 (1987)
\bibitem{MaisRipken} 
H. Mais, G. Ripken "Theory of spin-orbit motion in electron-positron
storage rings - summary of results" DESY 83-062  (1983)
\bibitem{Mais} 
H. Mais "Some topics in beam dynamics of storage rings"
 DESY 96-119  (1996)
\bibitem{Arnold} 
L. Arnold "Stochastische Differentialgleichungen"
R. Oldenbourg (1973)
\bibitem{vKampen}
N.G. van Kampen "Stochastic processes in physics and chemistry"
North Holland (1981)
\bibitem{Gardiner} 
C.W. Gardiner "Handbook of stochastic methods" Springer (1985)
\bibitem{Gard} 
T.C. Gard "Introduction to stochastic differential equations"
Marcel Dekker (1988)
\bibitem{Horsthemke} 
W. Horsthemke, R. Lefever "Noise induced transitions" Springer  (1984)
\bibitem{Kloeden} 
P.E. Kloeden, E. Platen  "Numerical solution of stochastic 
differential equations" Springer (1992)
\bibitem{Pauluhn} 
A. Pauluhn "Stochastic beam dynamics in storage rings" DESY-93-198
(1993)
\bibitem{Risken} 
H. Risken "The Fokker-Planck equation"  Springer (1989)
\bibitem{Richtm} 
R.D. Richtmyer, K.W. Morton "Difference methods for initial-value
problems" Interscience Publ. (1967)
\bibitem{Zorzano}
M.P. Zorzano "Numerical Integration of the Fokker-Planck
Equation and Application to Stochastic Beam Dynamics in
Storage Rings" PhD thesis to be submitted (1998) 
\bibitem{Kheifets}
S. Kheifets "Application of the Green's function method to some 
nonlinear problems of an electron storage ring, Part IV: Study
of a weak-beam interaction with a flat strong beam" Part. Accel.
\underline{15}, 153 (1984)
\bibitem{Gerasimov} 
A.L. Gerasimov "Phase convection and distribution "tails" in
periodically driven Brownian motion" Physica \underline{D41}, 89 (1990)
\bibitem{Izraelev}
F.M. Izraelev "Nearly linear mappings and their applications"
Physica \underline{D1}, 243 (1980)
\bibitem{www} 
see for example: "Benchmark Problem in Computational
Stochastic Dynamics: The Beam-Beam Problem in Accelerator Physics"
(http://www.nd.edu/$\sim$johnsone/beambeam/) 
\bibitem{Dome} 
G. D\^{o}me "Diffusion due to rf noise"
CERN Accelerator School, Oxford 1985, CERN 87-03 (1987)
\bibitem{Shih} 
H.J. Shih, J.A. Ellison, B. Newberger, R. Cogburn  "Longitudinal beam
dynamics with rf noise" Part. Accel. \underline{43}, 159 (1994)
\bibitem{Krinsky}
S. Krinsky, J.M. Wang "Bunch diffusion due to rf-noise" Part. Accel
\underline{12}, 107 (1982)
\bibitem{Sen} 
T. Sen, J.A. Ellison "HERA-B and halo control using noise"
DESY HERA 96-09 (1996)
\bibitem{Bazzani}
A. Bazzani, H. Mais "Effect of a coloured noise on the betatronic
motion: A possible mechanism for slow extraction" in Proc.
"Nonlinear and Stochastic Beam Dynamics in Accelerators - a
Challenge to Theoretical and Computational Physics",
L\"{u}neburg 1997, DESY 97-161
\bibitem{analogue}
D.G. Luchinsky, P.V.E. McClintock, M.I. Dykman "Analogue studies of
nonlinear systems" Rep. Prog. Phys. \underline{61}, 889 (1998)
\bibitem{Jung} 
P. Jung "Periodically driven stochastic systems" Phys. Rep. 
\underline{234}, 175 (1993)
\end{thebibliography}
\end{document}